\documentstyle[preprint,aps]{revtex}
\begin{document}
\draft
\title{Surface Critical Phenomena and Scaling\\ in the Eight-Vertex Model}
\author{M. T. Batchelor and Y. K. Zhou\cite{byline}}
\address{Department of Mathematics, School of Mathematical Sciences,
Australian National University, Canberra ACT 0200, Australia}
\date{September 1, 1995}
\maketitle
\begin{abstract}
We give a physical interpretation of the entries of the reflection
$K$-matrices of Baxter's eight-vertex model in terms of an Ising
interaction at an open boundary. Although the model still defies an
exact solution we nevertheless obtain the exact surface free energy
from a crossing-unitarity relation. The singular part of the surface
energy is described by the critical exponents
$\alpha_s = 2 - \frac{\pi}{2\mu}$ and $\alpha_1 = 1 - \frac{\pi}{\mu}$,
where $\mu$ controls the strength of the four-spin interaction.
These values reduce to the known Ising exponents
at the decoupling point $\mu=\pi/2$ and confirm the scaling relations
$\alpha_s = \alpha_b + \nu$ and $\alpha_1 = \alpha_b -1$.
\end{abstract}
\pacs{}

Our understanding of phase transitions and critical phenomena has
been greatly enhanced by the study of exactly solved lattice models
in statistical mechanics \cite{bbook}. Chief among these models is Baxter's
eight-vertex model, which exhibits continuously
varying critical exponents \cite{b1}.
Such exact results provide valuable insights into the key theoretical
developments of universality, renormalisation  and scaling.
The eight-vertex model is equivalent (see Figs. 1 and 2) to two Ising models
coupled together by four-spin interactions \cite{w,kw}.
{}From \cite{bbook,b1,w,kw} the singular part of the bulk free energy
$f_b$ scales as $f_b \sim |t|^{\pi/\mu}$ as $t \rightarrow 0$. Here
$t$ vanishes linearly with $T - T_c$, where $T_c$ is the critical
temperature. The variable
$\mu$ measures the strength of the four-spin interaction $M$ via
\begin{eqnarray}
\exp(2 M) = \tan (\mu/2).
\end{eqnarray}
When $\mu = \pi/m$, where $m$ is an even integer, the critical behaviour
is modified to $f_b \sim |t|^{\pi/\mu} \log |t|$.
This is the case in the Ising limit, where $\mu = \pi/2$.
The critical exponent describing the divergence of the bulk specific
heat, $C_b\sim |t|^{-\alpha_b}$ as $t \rightarrow 0$, is given by
$
\alpha_b = 2 - \pi/\mu,
$
with $\alpha_b = 0$ (logarithmic) for the Ising model.

A significant test of the scaling relations between critical exponents
was given by Johnson, Krinsky and McCoy \cite{jkm}, who derived the
correlation length exponent $\nu = \frac{\pi}{2 \mu}$ for
the eight-vertex model. Together with Baxter's result for $\alpha_b$
this confirmed the validity of the bulk scaling
law \cite{bbook,bin,diehl}
$
\alpha_b = 2 - 2 \nu.
$
However, the situation is not so satisfactory for the {\em surface} critical
behaviour \cite{bin,diehl}, as the eight-vertex model has not been solved
for {\em open} boundary conditions as in Fig.2.
Whereas integrability in the bulk is governed by solutions of the
Yang-Baxter equation \cite{mcg,y,b1},
integrability in the presence of boundaries is governed by solutions of
both the Yang-Baxter and reflection equations \cite{c,s}.
$K$-matrices satisfying the reflection equations have been found for the
eight-vertex model \cite{hy,cg,vg,ik},
but the diagonalisation of the transfer matrix remains a formidable problem.
Here we nevertheless derive two surface critical exponents of the
eight-vertex model, allowing a direct test of the proposed
scaling relations between bulk and surface critical exponents
\cite{bin,diehl,ws}.

The relation between the bulk Boltzmann weights $a,b,c,d$ of the
eight-vertex model
and the Ising couplings $K,L,M$ is depicted in Fig.1.
These weights are given by \cite{bbook}
\begin{eqnarray}
a(u)&=&\rho_0\, \theta_4(\lambda) \theta_4(u) \theta_1(\lambda-u),
\nonumber\\
b(u)&=&\rho_0\, \theta_4(\lambda) \theta_1(u) \theta_4(\lambda-u),
\nonumber\\
c(u)&=&\rho_0\, \theta_1(\lambda) \theta_4(u) \theta_4(\lambda-u),
\nonumber\\
d(u)&=&\rho_0\, \theta_1(\lambda) \theta_1(u) \theta_1(\lambda-u),
\label{abcd}
\end{eqnarray}
where $\rho_0$ is a normalization factor. Here $\theta_1(u)$ and $\theta_4(u)$
are the elliptic theta functions,
\begin{eqnarray}
\theta_1(u)&=& 2 \, q^{1/4} \sinh \frac{\pi u}{2 I}
    \prod_{n=1}^{\infty} \left(
  1-2 q^{2n} \cosh \frac{\pi u}{I}+q^{4n} \right) (1-q^{2n}), \nonumber \\
\theta_4(u)&=&\prod_{n=1}^{\infty} \left(
  1-2 q^{2n-1} \cosh \frac{\pi u}{I} +q^{4n-2}\right) (1-q^{2n}),\nonumber
\end{eqnarray}
of nome $q=\exp(-\pi I'/I)$, where $I$ and $I'$ are the half-period
magnitudes. In the principal regime, $0 < u < \lambda$, with
$0 < \lambda < I'$ and $0 < q < 1$.
Here $q \rightarrow 1$ at criticality.
In terms of the vertex weights, the bulk Ising couplings are given by
\begin{eqnarray}
\exp(4 K) &=& \frac{a c}{b d} =
\left[\frac{\theta_4(u)}{\theta_1(u)}\right]^2, \label{defk}\\
\exp(4 L) &=& \frac{a d}{b c} =
\left[\frac{\theta_1(\lambda-u)}{\theta_4(\lambda-u)}\right]^2, \\
\exp(4 M) &=& \frac{a b}{c d} =
\left[\frac{\theta_4(\lambda)}{\theta_1(\lambda)}\right]^2.
\end{eqnarray}
In the Ising limit, $M=0$ when $\lambda = \frac{1}{2} I'$, with the
spectral variable $u$ controlling the anisotropy of the Ising
couplings $K$ and $L$.

For the lattice orientation of Fig. 2, the
integrable boundary vertex weights can be written down
from the entries of the $K$-matrix. Now for the eight-vertex model
this reflection matrix is of the general form
$K^-(u) = K(u;\xi_-,\eta_-,\tau_-)$, with
\begin{eqnarray}
K(u;\xi,\eta,\tau)&=& \mbox{\small $\pmatrix{K_{11}(u)&K_{12}(u) \cr
                  K_{21}(u)&K_{22}(u)}$}
\end{eqnarray}
where \cite{ik}
\begin{eqnarray}
K_{11} &=& \rho_s\frac{\theta_1(\xi-u)}{\theta_4(\xi-u)},\\
K_{22} &=& \rho_s\frac{\theta_1(\xi+u)}{\theta_4(\xi+u)},\\
K_{12} &=&\rho_s\; \eta\; \theta_4^2(\xi) \frac{\theta_1(2u)}{\theta_4(2u)}
\frac{\left\{ \tau \left[ \theta_4^2(u) + \theta_1^2(u) \right] -
            \theta_1^2(u) + \theta_4^2(u) \right\}}
{\theta_4^2(\xi) \theta_4^2(u) - \theta_1^2(\xi) \theta_1^2(u)},\\
K_{21} &=&\rho_s\; \eta\; \theta_4^2(\xi) \frac{\theta_1(2u)}{\theta_4(2u)}
\frac{\left\{ \tau \left[ \theta_4^2(u) + \theta_1^2(u) \right] +
            \theta_1^2(u) - \theta_4^2(u) \right\}}
{\theta_4^2(\xi) \theta_4^2(u) - \theta_1^2(\xi) \theta_1^2(u)}.
\label{kel}
\end{eqnarray}
Here $\rho_s$ is a normalization factor and $\xi, \eta, \tau$ are
arbitrary parameters. In principle, these
three parameters are related to the surface couplings. We argue
that the variable $\xi$ controls the strength of the Ising surface
coupling $K_s$.
Similar to the bulk case, we see from
Fig. 1 that $\exp(4K_s)$ is given by a ratio of the boundary weights $r_{ij}$,
which in turn are related to the $K$-matrix elements \cite{yb},
\begin{eqnarray}
\exp(4K_s) = \frac{r_{11} r_{22}}{r_{12} r_{21}} =
\frac{K_{11}(u/2) K_{22}(u/2)} {K_{12}(u/2) K_{21}(u/2)}.\label{Ks}
\end{eqnarray}
The particular choice $\tau =0$ and $\xi = \frac{1}{2} I'$
simplifies to
\begin{eqnarray}
\exp(4K_s) = - \frac{1}{\eta^2}
\left[\frac{\theta_4(u)}{\theta_1(u)}\right]^2.
\end{eqnarray}
Comparison with the bulk parametrisation (\ref{defk}) implies that
the further choice of $\eta^2 = -1$ leads to $K=K_s$, i.e.
equal bulk and surface couplings in the Ising spin formulation.
These particular values of $\eta$ and $\tau$ can be chosen for all
$\xi$, since the surface coupling $K_s$ can be clearly set to be
independent of
$\eta$ and $\tau$, which can be seen from the product
$r_{11} r_{22}$.

The surface free energy can be obtained by applying the
inversion relation method,
which is known to give the correct
bulk free energy of the eight-vertex model
(see, e.g. \S 13.7 of Ref. \cite{bbook}).
By using the fusion procedure, the
transfer matrix of the eight-vertex model with boundaries described
by $K^\pm$-matrices has recently been found
to satisfy a group of functional relations \cite{z1,z2}.
Ignoring the finite-size corrections, which are of no
relevance here, the relations give the desired crossing-unitarity relation
for the transfer matrix eigenvalues \cite{foot1},
\begin{eqnarray}
\Lambda(u) \Lambda(u+\lambda)
&=&{\omega_+(u)\omega_-(u) }\rho^{2N}(u).
\label{inv-E}
\end{eqnarray}
The factor
\begin{eqnarray}
\rho(u)=\displaystyle{\theta_1(\lambda-u)\theta_1(\lambda+u)\over
\theta_1(\lambda)\theta_1(\lambda)}
\end{eqnarray}
is a bulk contribution whereas the product $\omega_+(u)\omega_-(u)$
is a surface contribution, with \cite{z1,z2}
\begin{eqnarray}
\omega_+(u)&=& K^+_{11}(u)b(-2u+\lambda)K^+_{22}(u+\lambda)
              +K^+_{12}(u)d(-2u+\lambda)K^+_{12}(u+\lambda) \nonumber \\
           && -K^+_{21}(u)a(-2u+\lambda)K^+_{12}(u+\lambda)
              -K^+_{22}(u)c(-2u+\lambda)K^+_{22}(u+\lambda),\\
\omega_-(u)&=&K^-_{21}(u+\lambda)d(2u+\lambda)K^-_{21}(u)
             +K^-_{22}(u+\lambda)b(2u+\lambda)K^-_{11}(u)\nonumber \\
           &&-K^-_{11}(u+\lambda)c(2u+\lambda)K^-_{11}(u)
             -K^-_{12}(u+\lambda)a(2u+\lambda)K^-_{21}(u).
\end{eqnarray}
Here $K^+(u)$ is the transpose of $K^-(-u+\lambda)$
with $\xi_-$ replaced by $\xi_+$, etc.

The bulk and surface free energies must both satisfy the
crossing-unitarity relation (\ref{inv-E}).
The surface energy can be separated from the bulk energy.
As we are only predominately interested here in the surface critical behaviour
rather than the precise form of the surface energy,
we consider only the diagonal elements of the $K$-matrix.
These terms are sufficient to extract the critical exponents
and physically we do not anticipate any change in the critical
behaviour arising from the off-diagonal terms \cite{foot2}.
Define $\Lambda_b = \kappa_b^{2 N}$ and $\Lambda_s = \kappa_s$, then
the bulk and surface free energies per site are defined by
$f_b(u)=-\log \kappa_b(u)$ and $f_s(u) = -\log \kappa_s(u)$.
{}From (13)-(16) we have
\begin{eqnarray}
\kappa_b(u)\kappa_b(u+\lambda)&=&\rho(u)
\label{inv-b}
\end{eqnarray}
for the bulk and
\begin{eqnarray}
\kappa_s(u)\kappa_s(u+\lambda)&=&{
   \theta_1(\xi_--u)\theta_1(\xi_-+u)
   \theta_1(\xi_+-u)\theta_1(\xi_++u)\over \theta_1^4(\lambda)}
{\theta_1(2\lambda-2u)\theta_1(2\lambda+2u)
    \over \theta_1^2(2\lambda)}\label{inv-s}
\end{eqnarray}
for the surface.

We obtain the solution of (\ref{inv-s}) for $\kappa_s(u)$ by
applying the inversion relation method \cite{bbook}. Let us first
recall the derivation of $\kappa_b(u)$ from (\ref{inv-b}).
It is convenient to introduce the variables
\begin{eqnarray}
x=\exp\left(-\pi\lambda/2 I\right)
\quad\mbox{and}\quad w=\exp\left(-\pi u/I\right).
\end{eqnarray}
To obtain $f_b(w)$ the argument is to assume that
$\kappa_b(w)$ is analytic and nonzero in the annulus $x^2\le w\le 1$,
allowing the Laurent expansion of $f_b(w)$,
\begin{eqnarray}
\log \kappa_b(w)=\sum_{n=-\infty}^{\infty} c_n w^n\;.
\end{eqnarray}
Inserting this expansion into the logarithm of both sides of (\ref{inv-b})
and equating the coefficients of powers of $w$ then gives
\begin{eqnarray}
f_b(w)=-\sum_{n=1}^{\infty}
{(x^{2n}+q^{2n}x^{-2n})(1-w^n)(1-x^{2n}w^{-n})\over
          n(1+x^{2n})(1-q^{2n})}\;. \label{bfree}
\end{eqnarray}
This is the desired result, from which the critical behaviour in the
limit $q \rightarrow 1$ is extracted by use of the
Poisson summation formula \cite{bbook}.
In terms of the variable $\mu = \pi \lambda/I'$, where
$I' \rightarrow \pi/2$ as $q \rightarrow 1$, it follows that
$f_b \sim p^{\pi/\mu}$ as $p \rightarrow 0$,
with $f_b \sim p^{\pi/\mu}\log p$ if ${\pi/\mu}$ is an even integer.
Here the conjugate nome $p=\exp(-2 \pi I/I')$ vanishes linearly
with the deviation from criticality variable $t$ \cite{bbook}.

We obtain the surface free energy by solving (\ref{inv-s}) under the
{\em same} analyticity assumptions as for the bulk case, together
with the further assumption that $\kappa_s(w)$ is analytic and
nonzero in the annulus $x<y_\pm<1$, where we have defined
$
y_\pm=\exp\left(-\pi \xi_\pm/ 2 I\right).
$
In this way we arrive at the result
\begin{eqnarray}
f_s(w,y_\pm)
&=&\sum_{n=1}^\infty
{(y_+^{2n}+q^{2n}y_+^{-2n}+y_-^{2n}+q^{2n}y_-^{-2n})(w^n+x^{2n}w^{-n})\over
          n(1+x^{2n})(1-q^{2n})} \nonumber \\
&&-\sum_{n=1}^\infty{(x^{4n}+q^{2n}x^{-4n})
   (1-w^{2n})(1-x^{4n}w^{-2n})\over   n(1+x^{4n})(1-q^{2n})}.
\label{sfree}
\end{eqnarray}
Applying the Poisson summation formula leads to a series for
$f_s$ in powers of the nome $p$.

The phenomenology of critical behaviour at a surface is well
developed \cite{bin,diehl,ws}.
In this case two surface critical exponents can be obtained
from the surface free energy; one
from the surface specific heat, $C_s \sim |t|^{-\alpha_s}$, and the
other from the ``local" specific heat in the boundary layer,
$C_1 \sim |t|^{-\alpha_1}$. Here the corresponding surface internal energy
is given by
\begin{eqnarray}
e_s(p)\sim {\partial f_s(u,\xi_\pm)\over\partial p} +e_1(p),
\end{eqnarray}
where $e_1(p)$ is the first layer internal energy,
\begin{eqnarray}
e_1(p)\sim{\partial f_s(u,\xi_\pm)\over\partial{\xi_\pm}}.
\end{eqnarray}
The related specific heats follow as
\begin{eqnarray}
 C_s \sim {\partial e_s\over\partial p}
\quad \mbox{and} \quad
 C_1 \sim {\partial e_1\over\partial p}.
\end{eqnarray}
These definitions follow from \cite{bin,diehl,ws} with the identifications
$p\sim t$ and $\xi_\pm\sim K_s$.
{}From (\ref{sfree}) we find that as $p\to 0$
\begin{eqnarray}
e_s(p)\sim p^{{\pi\over 2\mu}-1} \quad \mbox{and} \quad
e_1(p)\sim p^{\pi/\mu}.
\end{eqnarray}
As for the bulk case, a logarithmic factor appears if ${\pi/\mu}$ is an
even integer, with
\begin{eqnarray}
e_s(p)\sim p^{{\pi\over 2\mu}-1}\log p \quad \mbox{and} \quad
e_1(p)\sim p^{\pi/\mu}\log p.
\end{eqnarray}
This behaviour is to be compared at $\mu = \pi/2$ with the known Ising
results where the logarithmic factor is observed \cite{mw,ff,ay,r}.

In summary, we have derived the exact critical surface exponents
\begin{eqnarray}
\alpha_s=2-{\pi\over 2\mu} \quad \mbox{and} \quad
\alpha_1=1-{\pi\over\mu}
\end{eqnarray}
for the eight-vertex model.
At $\mu = \pi/2$, $\alpha_s=1$ (log) and $\alpha_1=-1$ (log),
in agreement with the Ising results \cite{mw,ff,ay,r}.
Recalling the bulk exponents $\alpha_b = 2 - {\pi\over\mu}$ \cite{b1}
and $\nu = {\pi\over 2\mu}$ \cite{jkm} we are thus able to provide
a significant confirmation of the scaling relations \cite{bin,diehl,ws}
\begin{eqnarray}
\alpha_s = \alpha_b + \nu \quad \mbox{and} \quad \alpha_1 = \alpha_b -1
\end{eqnarray}
between bulk and surface critical exponents.
The derivation of other surface exponents awaits the
diagonalisation of the transfer matrix, which remains a formidable open
problem.

We have found that the surface free energy scales
as $f_s \sim p^{\pi/2\mu}$ as $p\rightarrow 0$.
It is interesting to observe that this is in agreement with the
scaling behaviour of the interfacial tension \cite{bint}.
However, these two quantities differ away from criticality.

Finally we note that, in the same spirit as this work, the critical
magnetic surface exponent $\delta_s$ of the
two-dimensional Ising model
in a magnetic field \cite{bin,diehl} should also be obtainable from the
dilute $A_3$ model \cite{wns,roche}, which is known to lie in the same
universality class as the Ising model in a magnetic field \cite{wns}.
However, unlike for the eight-vertex model, where we have been able
to disentangle the critical behaviour from the known solution of the
reflection equations, the reflection matrices for the dilute $A_3$
model are yet to be determined.

\acknowledgments

The authors thank Professor R. J. Baxter for his interest and
encouragement in this work and the Australian Research Council
for support.

\begin{figure}

\def\oneup{\put( 56,776){\vector(1,1){0}}}
\def\twoup{\put( 64,776){\vector(-1,1){0}}}
\def\threeup{\put( 70,789){\vector(1,1){0}}}
\def\fourup{\put( 51,789){\vector(-1,1){0}}}
\def\onedown{\put( 50,770){\vector(-1,-1){0}}}
\def\twodown{\put( 70,770){\vector(1,-1){0}}}
\def\threedown{\put( 63,783){\vector(-1,-1){0}}}
\def\fourdown{\put( 57,783){\vector(1,-1){0}}}
\def\rightoneup{\put(116,716){\vector(1,1){0}}}
\def\righttwoup{\put(111,729){\vector(-1,1){0}}}
\def\rightonedown{\put(109,709){\vector(-1,-1){0}}}
\def\righttwodown{\put(116,724){\vector(1,-1){0}}}
\def\cross#1#2#3#4#5{
\setlength{\unitlength}{0.01000in}%
\begin{picture}(30,33)(45,776)
\put( 45,795){\line( 1,-1){ 30}}
\put( 75,795){\line(-1,-1){ 30}}
\put( 43,777){\tiny #1}
\put( 56,763){\tiny #2}
\put( 69,777){\tiny #3}
\put( 56,789){\tiny #4}{#5}
\end{picture}}
\def\rightK#1#2#3#4{
\setlength{\unitlength}{0.0110in}%
\begin{picture}(15,30)(105,713)
\put(105,735){\line( 1,-1){ 15}}
\put(120,720){\line(-1,-1){ 15}}
\put(102,717.5){\tiny #1}
\put(117,707){\tiny #2}
\put(117,728){\tiny #3}{#4}
\end{picture}}

\newcommand{\vs}[1]{\vspace*{#1cm}}
\newcommand{\hs}[1]{\hspace*{#1cm}}
\vskip 1cm
\hskip 3cm
\cross ++++{\oneup\twoup\threeup\fourup} \hs{0.4}
\cross {$-$}{$-$}{$-$}{$-$}{\oneup\twoup\threeup\fourup} \hs{0.4}
\cross +{$-$}+{$-$}{\onedown\twodown\threedown\fourdown} \hs{0.4}
\cross {$-$}+{$-$}+{\onedown\twodown\threedown\fourdown} \hs{0.4}
$a=A\;e^{K+L+M}$
\vskip 3mm
\hskip 3cm
\cross {$-$}{$-$}++{\oneup\twodown\threeup\fourdown} \hs{0.4}
\cross ++{$-$}{$-$}{\oneup\twodown\threeup\fourdown} \hs{0.4}
\cross +{$-$}{$-$}+{\onedown\twoup\threedown\fourup} \hs{0.4}
\cross {$-$}++{$-$}{\onedown\twoup\threedown\fourup} \hs{0.4}
$b=A\;e^{-K-L+M}$
\vskip 3mm
\hskip 3cm
\cross ++{$-$}+{\oneup\twodown\threedown\fourup} \hs{0.4}
\cross {$-$}{$-$}+{$-$}{\oneup\twodown\threedown\fourup} \hs{0.4}
\cross {$-$}+++{\onedown\twoup\threeup\fourdown} \hs{0.4}
\cross +{$-$}{$-$}{$-$}{\onedown\twoup\threeup\fourdown} \hs{0.4}
$c=A\;e^{K-L-M}$
\vskip 3mm
\hskip 3cm
\cross {$-$}{$-$}{$-$}+{\oneup\twoup\threedown\fourdown} \hs{0.4}
\cross +++{$-$}{\oneup\twoup\threedown\fourdown} \hs{0.4}
\cross +{$-$}++{\onedown\twodown\threeup\fourup} \hs{0.4}
\cross {$-$}+{$-$}{$-$}{\onedown\twodown\threeup\fourup} \hs{0.4}\vs{1}
$d=A\;e^{-K+L-M}$\
\vskip 3mm
\hskip 5cm
\rightK +++{\rightoneup\righttwoup}\hs{0.4}
\rightK {$-$}{$-$}{$-$}{\rightoneup\righttwoup}\hs{0.4}
$r_{11}=B\;e^{K_s+M^1_s+M^2_s}$
\vskip 3mm
\hskip 5cm
\rightK +{$-$}{$-$}{\rightonedown\righttwodown}\hs{0.4}
\rightK {$-$}++{\rightonedown\righttwodown}\hs{0.4}
$r_{22}=B\;e^{K_s-M^1_s-M^2_s}$
\vskip 3mm
\hskip 5cm
\rightK +{$-$}+{\rightonedown\righttwoup}\hs{0.4}
\rightK {$-$}+{$-$}{\rightonedown\righttwoup}\hs{0.4}
$r_{12}=B\;e^{-K_s+M^1_s-M^2_s}$
\vskip 3mm
\hskip 5cm
\rightK ++{$-$}{\rightoneup\righttwodown}\hs{0.4}
\rightK {$-$}{$-$}+{\rightoneup\righttwodown}\hs{0.4}\vs{1}
$r_{21}=B\;e^{-K_s-M^1_s+M^2_s}$
\vskip 1cm

\caption{The bulk and surface vertex and Ising spin configurations and
their corresponding Boltzmann weights. The nearest-neighbour bulk
interactions $K$ and $L$ are in the vertical and horizontal directions,
respectively. The four-spin interaction is denoted by $M$ and the general
nearest-neighbour surface interactions by $K_s, M_s^1$ and $M_s^2$
in the vertical, SW-NE and SE-NW directions, respectively.
The constants $A$ and $B$ do not enter into the critical properties.}
\end{figure}
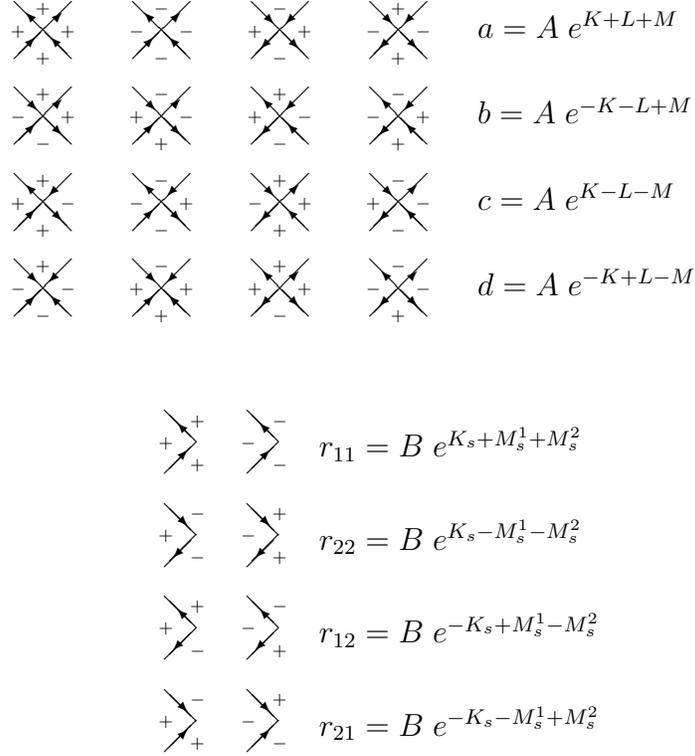

\newpage
\begin{center}
\begin{figure}
\vskip 1cm
\setlength{\unitlength}{0.01in}%
\begin{picture}(300,240)(105,540)
\multiput(105,780)(60,0){6}{\circle{6}}
\multiput(105,720)(60,0){6}{\circle{6}}
\multiput(105,660)(60,0){6}{\circle{6}}
\multiput(105,600)(60,0){6}{\circle{6}}
\multiput(135,750)(60,0){5}{\circle*{6}}
\multiput(135,690)(60,0){5}{\circle*{6}}
\multiput(135,630)(60,0){5}{\circle*{6}}
\multiput(135,801)(60,0){5}{\line( 0,-1){220}}
\multiput(135,750)(0,-60){3}{\line( 1, 0){240}}
\multiput(105,780)(9.1,0){33}{\line( 1, 0){8}}
\multiput(105,660)(9.1,0){33}{\line( 1, 0){8}}
\multiput(105,720)(9.1,0){33}{\line( 1, 0){8}}
\multiput(105,600)(9.1,0){33}{\line( 1, 0){8}}
\multiput(105,798)(0,-9.1){24}{\line( 0,-1){6}}
\multiput(405,798)(0,-9.1){24}{\line( 0,-1){6}}
\multiput(165,798)(0,-9.1){24}{\line( 0,-1){6}}
\multiput(225,798)(0,-9.1){24}{\line( 0,-1){6}}
\multiput(285,798)(0,-9.1){24}{\line( 0,-1){6}}
\multiput(345,798)(0,-9.1){24}{\line( 0,-1){6}}
\put(411,744){\small$K_s$}
\put(267,687){\small$L$}
\put(345,672){\small$K$}
\put(252,702){\small$K$}
\put(327,657){\small$L$}
\multiput(104,750)(6.36364,-6.36364){27}{$.$}
\multiput(104,750)(6.36364,6.36364){8}{$.$}
\multiput(104,690)(6.66667,6.66667){17}{$.$}
\multiput(104,690)(6.66667,-6.66667){16}{$.$}
\multiput(104,630)(6.42857,-6.42857){8}{$.$}%
\multiput(104,630)(6.42857,6.42857){27}{$.$}%
\multiput(374,600)(6.00000,6.00000){6}{$.$}
\multiput(374,600)(-6.00000,-6.00000){3}{$.$}
\multiput(404,630)(-6.31579,6.31579){27}{$.$}
\multiput(404,690)(-6.31579,6.31579){18}{$.$}
\multiput(404,690)(-6.31579,-6.31579){18}{$.$}
\multiput(404,750)(-6.31579,-6.31579){27}{$.$}
\multiput(344,750)(-6.31579,-6.31579){27}{$.$}
\multiput(344,750)(6.31579,6.31579){8}{$.$}
\multiput(284,750)(-6.31579,-6.31579){27}{$.$}
\multiput(284,750)(6.31579,6.31579){8}{$.$}
\multiput(404,750)(-6.31579,6.31579){8}{$.$}
\multiput(164,750)(6.36364,-6.36364){27}{$.$}
\multiput(224,750)(6.36364,-6.36364){27}{$.$}
\multiput(224,750)(-6.36364,6.36364){8}{$.$}
\multiput(164,750)(-6.36364,6.36364){8}{$.$}
\end{picture}

\caption{The geometric relation between the eight-vertex model lattice
(dotted lines) and the Ising model lattice (broken and solid lines).
The Ising lattice is divided into two sub-lattices (solid and open circles).}

\end{figure}
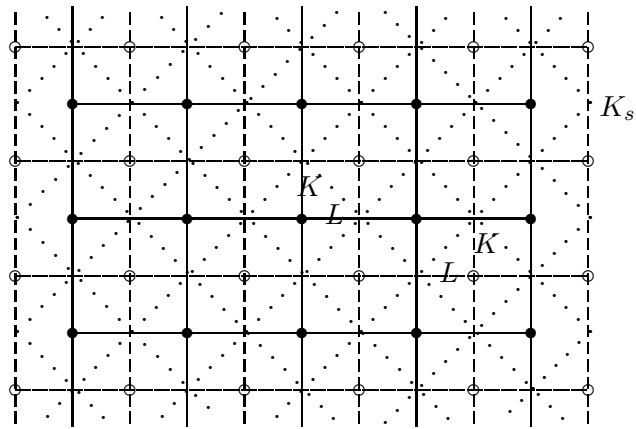
\end{center}


\begin{references}

\bibitem[*]{byline} On leave of absence from Institute of Modern
Physics, Northwest University, Xian 710069, China.
%
\bibitem{bbook} R. J. Baxter, {\it Exactly Solved Models in Statistical
Mechanics} (Academic, London, 1982).
%
\bibitem{b1} R. J. Baxter, Phys. Rev. Lett. {\bf 26}, 832 (1971);
Ann. Phys. (N.Y.) {\bf 70}, 193 (1972).
%
\bibitem{w} F. Y. Wu, Phys. Rev. B {\bf 4}, 2312 (1971).
%
\bibitem{kw} L. P. Kadanoff and F. J. Wegner,
Phys. Rev. B {\bf 4}, 3989 (1971).
%
\bibitem{jkm} J. D. Johnson, S. Krinsky and B. M. McCoy,
Phys. Rev. Lett. {\bf 29}, 492 (1972); Phys. Rev. A {\bf 8}, 2526 (1973).
%
\bibitem{bin} K. Binder, in {\it Phase Transitions and Critical Phenomena},
edited by C. Domb and J. L. Lebowitz, (Academic, London, 1983), Vol. 8, p 1.
%
\bibitem{diehl} H. W. Diehl, in {\it Phase Transitions and Critical Phenomena},
edited by C. Domb and J. L. Lebowitz, (Academic, London, 1986), Vol. 10, p 75.
%
\bibitem{mcg} J. B. McGuire, J. Math. Phys. {\bf 5}, 622 (1964).
%
\bibitem{y} C. N. Yang, Phys. Rev. Lett. {\bf 19}, 1312 (1967);
Phys. Rev. {\bf 168}, 1920 (1968).
%
\bibitem{c} I. V. Cherednik, Theor. Math. Phys. {\bf 61}, 977 (1984).
%
\bibitem{s} E. K. Sklyanin, J. Phys. A {\bf 21}, 2375 (1988).
%
\bibitem{hy} B. Y. Hou and R. H. Yue, Phys. Lett. A {\bf 183}, 169 (1993);
B. Y. Hou, K. J. Shi, H. Fan and Z. X. Yang,
Commun. Theor. Phys. {\bf 23}, 163 (1995).
%
\bibitem{cg} R. Cuerno and A. Gonz\'alez-Ruiz,
J. Phys. A {\bf 26}, L605 (1993).
%
\bibitem{vg} H. J. de Vega and A. Gonz\'alez-Ruiz,
J. Phys. A {\bf 27}, 6129 (1994).
%
\bibitem{ik} T. Inami and H. Konno, J. Phys. A {\bf 27}, L913 (1994).
%
\bibitem{ws}M. Wortis and N. M. ${\check {\rm S}}$vraki{$\acute {\rm c}$},
 IEEE Trans.   Mag. {\bf 18}, 721 (1982).
%
\bibitem{yb} C. M. Yung and M. T. Batchelor,
Nucl. Phys. B {\bf 435}, 430 (1995).
%
\bibitem{z1} Y. K. Zhou, Nucl. Phys. B {\bf 453}, 619 (1985).
%
\bibitem{z2} Y. K. Zhou,
{\sl Row transfer matrix functional
relations for Baxter's eight-vertex and six-vertex models with open
boundaries via more general reflection matrices}, hep-th/9510095,
Nucl. Phys. B (in press).
%
\bibitem{foot1} Crossing and unitarity relations
for boundary $K$-matrices have already been argued in terms of field
theory with a boundary, see
S. Ghoshal and A. Zamolodchikov,
Int. J. Mod. Phys. A {\bf 21}, 3841 (1994).
%
However, it is not clear how to write down
the boundary crossing and unitarity relations for an off-critical model
according to this theory. The transfer matrix fusion procedure
gives a consistent way of finding the crossing-unitarity relation.
%
\bibitem{foot2} The precise form of the surface free energy for the
geometry depicted in Fig. 2 can be obtained
by considering all terms and introducing alternating inhomogeneities
as in Ref. \cite{yb}. Full details of this calculation will be given
elsewhere.
%
\bibitem{mw} B. M. McCoy and T. T. Wu, Phys. Rev. {\bf 162}, 436 (1967).
%
\bibitem{ff} M. E. Fisher and A. E. Ferdinand,
Phys. Rev. Lett. {\bf 19}, 169 (1967).
%
\bibitem{ay} H. Au Yang, J. Math. Phys. {\bf 14}, 937 (1973).
%
\bibitem{r} P. Reed, J. Phys. A {\bf 11}, 137 (1978).
%
\bibitem{bint} R. J. Baxter, J. Stat. Phys. {\bf 8}, 25 (1973).
%
\bibitem{wns} S. O. Warnaar, B. Nienhuis and K. A. Seaton,
Phys. Rev. Lett. {\bf 69}, 710 (1992);
Int. J. Mod. Phys. {\bf 7}, 3727 (1993).
%
\bibitem{roche} Ph. Roche, Phys. Lett. B {\bf 285}, 49 (1992).
%
\end{references}
\end{document}